\documentclass[aps,prb,floatfix,twocolumn,superscriptaddress,showpacs,amsmath,amssymb]{revtex4-1}
\usepackage{graphicx}
\usepackage{epstopdf}
\usepackage{epsfig} 
\allowdisplaybreaks

\newcommand{\be}{\begin{equation}}
\newcommand{\ee}{\end{equation}}
\newcommand{\bea}{\begin{eqnarray}}
\newcommand{\eea}{\end{eqnarray}}
\newcommand{\ba}{\begin{eqnarray*}}
\newcommand{\ea}{\end{eqnarray*}}
\newcommand{\dagga}{{\phantom{\dagger}}}

\newcommand{\bq}{\mathbf{q}}

\newcommand{\br}{\mathbf{r}}
\newcommand{\bnot}{\mathbf{0}}

\newcommand{\dis}{\displaystyle}

\newcommand{\up}{\uparrow}
\newcommand{\down}{\downarrow}
\newcommand{\fract}[2]{\frac{\dis #1}{\dis #2}}

\newcommand{\eqn}[1]{(\ref{#1})}

\begin{document}

\title{Dynamical quantum phase transitions and broken-symmetry edges in the many-body eigenvalue spectrum}

\author{Giacomo Mazza} 
\affiliation{International School for
  Advanced Studies (SISSA), and CNR-IOM Democritos, Via Bonomea
  265, I-34136 Trieste, Italy} 
\author{Michele Fabrizio} 
\affiliation{International School for
  Advanced Studies (SISSA), and CNR-IOM Democritos, Via Bonomea
  265, I-34136 Trieste, Italy} 
\affiliation{The Abdus Salam
  International Centre for Theoretical Physics (ICTP), P.O.Box 586,
  I-34014 Trieste, Italy} 

\date{\today} 

\pacs{71.10.Fd, 05.30.Rt, 05.70.Ln}

\begin{abstract}
Many body models undergoing 
a quantum phase transition to a broken-symmetry phase that survives up to a critical temperature must possess, 
in the ordered phase, symmetric as well as non-symmetric eigenstates. 
We predict, and explicitly show 
in the fully-connected Ising model in a transverse field, that these two classes of eigenstates do not overlap in energy, 
and therefore that an energy edge exists
separating low-energy symmetry-breaking eigenstates from high-energy symmetry-invariant ones. This energy 
is actually responsible, as we show, for 
the dynamical phase transition displayed 
by this model under a sudden large increase of the transverse field.
A second situation we consider is the opposite, 
where the symmetry-breaking eigenstates are those in the high-energy sector of the spectrum, whereas the low-energy eigenstates are symmetric. 
In that case too a special energy must exist marking 
the boundary and leading to  
unexpected out-of-equilibrium dynamical behavior.  An 
example is the fermonic repulsive Hubbard model Hamiltonian 
$\mathcal{H}$. 
Exploiting the trivial fact that the high energy spectrum of 
$\mathcal{H}$ is also the low energy one of $-\mathcal{H}$, 
we conclude that the high energy eigenstates of the Hubbard model are superfluid. Simulating in a time-dependent 
Gutzwiller approximation the time evolution of a high energy BCS-like trial wave function, we show that a small superconducting order parameter will actually grow in spite of the repulsive nature of interaction. 

\end{abstract}
\maketitle

\section{Introduction}
The temporal quantum evolution of an isolated macroscopic system initially prepared in an out-of-equilibrium configuration is currently  turning from an abstract concept, useful for discussing fundaments of quantum statistical mechanics, to a real phenomenon that can be observed and studied experimentally. This metamorphosis has been mainly driven by experiments on cold atoms,\cite{BlochRMP,SilvaRMP} 
but it will be surely given further impulse in the near future by the fast progresses in time-resolved spectroscopy on condensed-matter systems. In fact, the early time dynamics (up to $\sim 1$ ns) of a  material that is driven out-of-equilibrium e.g. by an intense ultra-short laser pulse is still uninfluenced by the environmental heat sink, hence it is to a good approximation the dynamics of an isolated system.       

An interesting class of experiments focuses on the dynamics across phase transitions. In cold atom systems, this can be achieved by a sudden change of the experimental conditions, e.g. the depth of the optical lattice, which corresponds to suddenly altering the Hamiltonian parameters, a unique opportunity offered by these systems.\cite{Dynamics-1}  
In pump-probe experiments on real materials, one can instead tune the  fluence of the pumping laser, i.e. the excess energy injected into the system.\cite{Nasu} That energy is supposed to undergo fast redistribution among all degrees of freedom, first among the electrons (faster), later 
among the phonons (slower), thus effectively heating the sample 
and raising its temperature.
If the equilibrium phase diagram has a transition between a low temperature phase and a high temperature one, the effective temperature rise could drive such a phase transition, though the system will eventually relax back to its initial equilibrium state by the coupling to the external thermostat. 
In reality, a dynamical phase transition in an isolated macroscopic system is not as trivial as one could imagine. Across a thermodynamic transition, ergodicity is either lost or recovered, hence it is not at all obvious that the unitary time-evolution of an initial non-equilibrium quantum state should bring about the same results as, at equilibrium, the adiabatic change of a coupling constant or of temperature. 

This is actually a central question at the basis of quantum statistical mechanics. Here, more modestly, we highlight a possible link between the ergodicity breakdown at a phase transition and the occurrence in the many-body eigenvalue spectrum of "broken-symmetry edges", namely of 
special energies that mark the boundaries between symmetry-breaking and symmetry-invariant eigenstates. 
\begin{figure}[thb]
\vspace{0.2cm}
\centerline{\includegraphics[width=6cm]{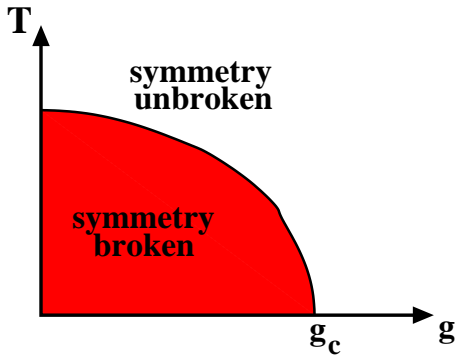}}
\caption{(Color online) Generic phase diagram of a model that 
possesses a symmetry broken phase in the temperature $T$ vs. coupling constant $g$ phase diagram, separated from a symmetric phase by a second order phase transition.} 
\label{Z2}
\end{figure}

Let us imagine a system described by a Hamiltonian $\mathcal{H}$, which undergoes a quantum phase transition, the zero-temperature endpoint of a whole second-order critical line that separates a low-temperature broken-symmetry phase from a high-temperature symmetric one, see Fig.~\ref{Z2} where $g$ is the coupling constant that drives the phase transition. 
For the sake of simplicity, and also because it will be relevant later, let us assume that the broken symmetry is a discrete $Z_2$ - all arguments below do not depend on this specific choice - with order parameter 
\be
\langle \sigma \rangle = 
\frac{1}{V}\sum_i\,\langle \sigma_i\rangle\in [-1,1],\label{OP}
\ee
which is not a conserved quantity, i.e. 
$\Big[\sigma,\mathcal{H}\Big]\not =0$, and where
$V\to\infty$ is the volume and $i$ labels lattice sites. 
Below the quantum critical point $g<g_c$, Fig. \ref{Z2}, 
the ground state is doubly degenerate and not $Z_2$ invariant. If $\mid \Psi_{\pm}\rangle$ are the two ground states, they can be chosen such that 
\[
\langle \Psi_{\pm}\mid\sigma\mid\Psi_{\pm}\rangle = \pm m,\qquad
\text{with } m>0.
\]
On the contrary, for $g>g_c$, the ground state is unique and symmetric, 
i.e. the average of $\sigma$ vanishes. Since the symmetry breaking survives at finite temperature, see Fig. \ref{Z2}, one must 
conclude that, besides the ground state, a whole macroscopic set of low energy states is $Z_2$ not-invariant. The ergodicity breakdown in a symmetry broken phase specifically implies that these states, in the example we are dealing with, are grouped into two subspaces that are mutually orthogonal in the thermodynamic limit, one that can be chosen to include all eigenstates with $\langle \sigma\rangle >0$, the other those with $\langle \sigma\rangle <0$. Since the symmetry is recovered above a critical temperature, then there should exist a high energy subspace that includes symmetry invariant eigenstates. We argue that there should be a special energy in the spectrum, a {\sl "broken-symmetry edge"} 
$E_*$, such that all eigenstates with $E<E_*$ break the symmetry, while all eigenstates above $E_*$ are symmetric. 
In the case where the Hamiltonian has additional symmetries besides $Z_2$, hence conserved quantities apart from energy, we claim that, within each subspace invariant under these further symmetries, there must exist an edge above which symmetry is restored, even though its value may differ from one subspace to another as is the case in the model discussed in the next section.
   
Let us for instance focus on any of these subspaces. The $Z_2$ symmetry 
implies that all eigenstates are even or odd under $Z_2$. 
The order parameter $\sigma$ is odd, hence its average value is strictly zero on any eigenstate, either odd or even. 
Nevertheless, we can formally define 
an order parameter $m(\Psi_E)$ of a given eigenstate $\mid\Psi_E\rangle$ through the positive square root of 
\be
m\big(\Psi_E\big) = \sqrt{\lim_{|i-j|\to\infty} 
\left|\langle \Psi_E\mid \sigma_i\sigma_j
\mid \Psi_E\rangle\right|}.\label{OP-1}
\ee
We denote by 
\be
\rho_{SB}(E) = \text{e}^{V\,S_{SB}(\epsilon)},
\ee
the density of symmetry-breaking eigenstates, namely those with 
$m(\Psi_E)>m_0$, where $m_0$ is a cut-off value that vanishes sufficiently fast as $V\to\infty$, being $\epsilon=E/V$ and $S_{SB}(\epsilon)$ their energy and entropy per unit volume. Seemingly, we define  
\be
\rho_{SI}(E)=\text{e}^{V\,S_{SI}(\epsilon)},
\ee
the density of the symmetry-invariant eigenstates, $m(\Psi_E)\leq m_0$, with $S_{SI}(\epsilon)$ their entropy. We claim that there exists an energy $E_*=V\epsilon_*$ that marks the microcanonical continuous phase transition in that specific invariant subspace, such that 
\bea
\lim_{V\to\infty} S_{SI}(\epsilon) &=& 0,\qquad \text{for }\epsilon<\epsilon_*, \label{S_SI}\\
\lim_{V\to\infty} S_{SB}(\epsilon) &=& 0,\qquad \text{for }\epsilon>\epsilon_*. \label{S_SB}
\eea
We do not have a rigorous proof of the above statement, but just a plausible argument. Let us suppose to define the average $m(E)>m_0$ 
of the order parameter over the symmetry-breaking 
eigenstates through
\be
m(\epsilon) = \fract{1}{\rho_{SB}(E)}\sum_{\Psi_{E'}} \,
m\big(\Psi_{E'}\big)\,\delta\big(E'-E\big).\label{m-SB}
\ee 
The actual microcanonical average is thus 
\be
\overline{m}(\epsilon) = 
\fract{\rho_{SB}(E)}{\rho_{SB}(E)+\rho_{SI}(E)}\,
m(\epsilon).\label{m-ave}
\ee 
In the thermodynamic limit $V\to\infty$, hence $m_0\to 0$, the continuous phase transition would imply the existence of an energy $\epsilon_*$ 
such that, for $\epsilon\lesssim \epsilon_*$, $\overline{m}(\epsilon)
\sim (\epsilon_*-\epsilon)^{\beta'}$, where the exponent $\beta'$ may not coincide with the corresponding one in the canonical ensemble $\overline{m}(T) \sim (T_c-T)^\beta$,
while $\overline{m}(\epsilon>\epsilon_*)=0$. Since the entropy ratio on the r.h.s. of Eq. \eqn{m-ave}
is either 1 or 0 in the thermodynamic limit, we conclude that the critical behavior comes from 
$m(\epsilon\lesssim\epsilon_*)\sim (\epsilon_*-\epsilon)^{\beta'}$, which, by continuity, implies $m(\epsilon>\epsilon_*)=0$, namely that there are no symmetry-breaking eigenstates with finite entropy density above $\epsilon_*$, hence Eq. \eqn{S_SB}. This further suggests that symmetry-breaking and symmetry-invariant eigenstates exchange their role across the transition, which makes also Eq. \eqn{S_SI} plausible. 
We do not exclude that symmetry-breaking eigenstates may survive above $\epsilon_*$, or vice versa for symmetric ones; we just state that, if they survive, their entropy is not extensive. 

We may also guess a generalization of the above picture to the most common situation of a first order phase transition. In this case we expect two different edges, $\epsilon_{1}<\epsilon_{*}$. Below $\epsilon_{1}$ the entropy density of symmetry-invariant states $S_{SI}(\epsilon)$ vanishes in the thermodynamic limit, while above $\epsilon_{*}$, the actual edge for symmetry restoration, it is $S_{SB}(\epsilon)$ that goes to zero.  

If we accept the existence of such an energy threshold, then we are also able to justify, without invoking any thermalization hypothesis,\cite{thermalization-1,thermalization-2} why a material, whose equilibrium phase diagram is like that of Fig. \ref{Z2}, may undergo a dynamical phase transition once supplied initially with enough excess energy so as to push it above $E_*$. 

We mention once more that the above arguments are not at all a real proof. However, they can be explicitly proven in mean-field like models, like the fully connected Ising model that we discuss in section \ref{Ising}. 
There, we explicitly demonstrate that the dynamical transition occurs because above a threshold energy there are simply no more broken-symmetry eigenstates in the spectrum. We believe this is important because it may happen that such an energy threshold, hence such a dynamical transition, exists also in models whose phase diagram is different from that of Fig. \ref{Z2}, as we are going to discuss in section  
\ref{Hubbard}.

\section{First model: the fully connected Ising in a transverse field}
\label{Ising}

We consider the Hamiltonian of an Ising model in a transverse field
\bea
\mathcal{H} &=& -\sum_{i,j}\,J_{ij}\,\sigma^z_i \sigma^z_j - h \sum_i\,\sigma^x_i\nonumber\\
&=& -\frac{1}{N}\sum_\bq\, J_\bq\,\sigma^z_\bq \sigma^z_{-\bq} 
- h\,\sigma^x_{\bnot},\label{Ham-0}
\eea
where $N$ is the number of sites,
\[
\sigma^a_\bq = \sum_i\,\text{e}^{-i\bq\cdot\br_i}\,\sigma^a_i,
\]
is the Fourier transform of the spin operators, and $J_\bq$ the Fourier transform of the exchange.
In the (mean-field) fully-connected limit, $J_\bq = J\,\delta_{\bq\bnot}$, the model
\eqn{Ham-0} simplifies into 
\be
\mathcal{H} = -\frac{1}{N}\,\sigma^z_\bnot \sigma^z_\bnot - h\,\sigma^x_\bnot
= -\frac{4}{N} S^z S^z - 2h\,S^x,\label{Ham}
\ee
having set $J=1$ and defined the total spin $\mathbf{S}=\boldsymbol{\sigma}_\bnot/2$.  
It turns out that the Hamiltonian Eq.~\eqn{Ham} can be solved exactly. We shall closely follow the work by Bapst and Semerjian,\cite{Semerjian} whose approach fits well our purposes. 
For reader's convenience we will repeat part of Bapst and Semerjian's calculations.   
We start by observing that the Hamiltonian 
\eqn{Ham} commutes with the total spin operator $\mathbf{S}\cdot\mathbf{S}$, with eigenvalue 
$S(S+1)$, so that one can diagonalize $\mathcal{H}$ within each $S\in [0,N/2]$ sector, which contains 
$2S+1$ distinct eigenvalues, each one $g(S)$ times degenerate, where $g(N/2)=1$ and, for $S<N/2$,  
\be
g(S)=\binom{N}{\fract{N}{2}+S} - \binom{N}{\fract{N}{2}+S+1},
\label{g(S)}
\ee
which is the number of ways to couple $N$ spin-1/2 to obtain total spin $S$.
We define 
\be
S =  N\Big(\frac{1}{2} - k\Big) ,\label{p}
\ee
where $k$, for large $N$, becomes a continuous variable $k\in [0,1/2]$. For a given $S$, a generic eigenfunction can be written as 
\be
\mid \Phi_E\rangle = \sum_{M=-S}^S\,\Phi_E(M)\mid M\rangle,\label{Psi}
\ee
where $\mid M\rangle$ is eigenstate of $S^z$ with eigenvalue $M\in [-S,S]$. 
One readily find the eigenvalue equation\cite{Semerjian}
\bea
E\,\Phi_E(M) &=& - \frac{4}{N}\,M^2\,\Phi_E(M) \label{eigen-1}\\
&& - h\bigg[
\sqrt{S(S+1) - M(M-1)}\;\Phi_E(M-1) \nonumber\\
&& ~~~+ \sqrt{S(S+1) - M(M+1)}\;\Phi_E(M+1)\bigg].\nonumber
\eea
We now assume $N$ large keeping $k$ constant. We also define 
\[
m = \frac{2M}{N} \in [-1+2k,1-2k], 
\]
so that, at leading order in $N$, after setting $E=N\epsilon$ and 
\[
\Phi_E(M) = \Phi_\epsilon(m),
\]
the Eq.~\eqn{eigen-1} reads
\bea
\epsilon\,\Phi_\epsilon(m) &=& - m^2\,\Phi_\epsilon(m) 
- \frac{h}{2}\,\sqrt{(1-2k)^2-m^2}\nonumber\\
&& \;\Bigg[\Phi_\epsilon\left(m-\frac{2}{N}\right) + 
\Phi_\epsilon\left(m+\frac{2}{N}\right)\Bigg].\label{eigen-2}
\eea
Following Ref.\onlinecite{Semerjian}, we set 
\be
\Phi_\epsilon(m) \propto \exp\big[ -N\,\phi_\epsilon(m)\big],\label{ansatz}
\ee
where the proportionality constant is the normalization, so that 
\ba
\Phi_\epsilon\left(m\pm \frac{2}{N}\right) &\propto& 
\exp\bigg[ -N\,\phi_\epsilon\left(m\pm \frac{2}{N}\right)\bigg]\\
&& \simeq \Phi_\epsilon(m)\,\text{e}^{\mp 2\,\phi'_\epsilon(m)},
\ea
Upon substituting the above expression into \eqn{eigen-2}, the following equation follows
\be
\phi'_\epsilon(m) = 
\frac{1}{2}\,\text{arg}\,\cosh\bigg(- \fract{\epsilon+m^2}{h\sqrt{(1-2k)^2-m^2}}\bigg).\label{eigen-3}
\ee

\begin{figure}[t]
\centerline{\includegraphics[width=7cm]{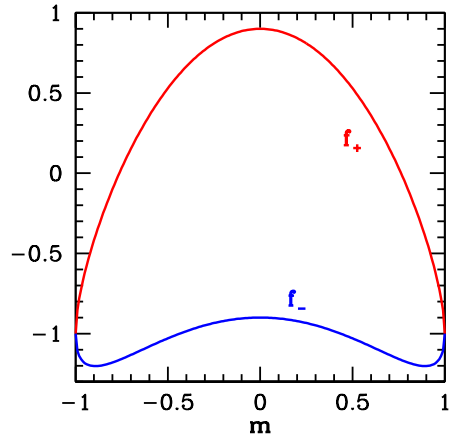}}
\vspace{-0.8cm}
\caption{\label{1} The two function $f_+(m)$ and $f_-(m)$ for $k=0$ and 
$h=0.9$. The allowed values of the eigenvalues are those 
between the minimum of $f_-$ and the maximum of $f_+$.}
\end{figure}

\begin{figure}[bht]
\centerline{\includegraphics[width=7cm]{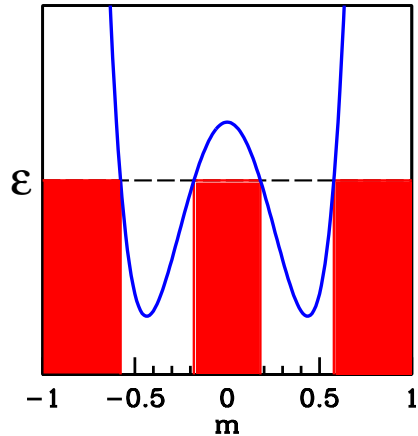}}
\vspace{-0.8cm}
\caption{\label{evan} (Color online) For a given energy $\epsilon<\epsilon_*$, we draw in red the regions $\epsilon<f_-(m)$ where the wave function vanishes exponentially as $N\to\infty$.}
\end{figure}

For large $N$, in order for the wave function \eqn{ansatz} to be normalizable, we must impose that: 
({\sl i}) the $\Re \text{e} \,\phi_\epsilon(m) \geq 0$; ({\sl ii}) the $\Re \text{e} \,\phi_\epsilon(m)$ must have zeros, which, because of ({\sl i}), are also minima. As showed in Ref. \onlinecite{Semerjian}, these two 
conditions imply that the allowed values of the energy are 
\be
\text{min}\Big(f_-(m)\Big)\leq \epsilon \leq \text{Max}\Big(f_+(m)\Big),\label{c-1},
\ee
where 
\bea
f_+(m) &=& - m^2  + h\,\sqrt{(1-2k)^2-m^2},\label{f+}\\
f_-(m) &=&  - m^2  - h\,\sqrt{(1-2k)^2-m^2} .\label{f-}
\eea
At fixed $k$, the lowest allowed energy is thus  
\be
\epsilon_\text{min}(k)=\text{min}\left(f_-(m)\right)  = - (1-2k)^2 - \frac{h^2}{4},\label{e<-eq}
\ee
and occurs at 
\be
m^2(k) = (1-2k)^2 - \frac{h^2}{4},\label{m-eq}
\ee
if $h\leq h(k) = 2(1-2k)$, otherwise the minimum energy occurs at $m=0$, 
\be
\epsilon_\text{min}(k) = f_-(0) = -h\,(1-2k).\label{e>-eq}
\ee
It follows that the actual ground state is always in the $k=0$ subspace and has energy 
\be
\epsilon_0 = \begin{cases}
-1 - \frac{h^2}{4} & \mbox{if } h\leq h(0)=2,\\
-h & \mbox{if } h>h(0).
\end{cases}
\ee
In Fig. \ref{1} we plot the two functions $f_+(m)$ and $f_-(m)$ for $k=0$ and $h=0.9<h(0)$. 
As shown by Bapst and Semerjian,\cite{Semerjian} whenever $f_-(m)$ has a double minimum 
as in Fig.~\ref{1}, any eigenstate with energy below $\epsilon <
\epsilon_*=f_-(m=0)$
is doubly degenerate in the thermodynamic limit $N\to\infty$, being localized either at positive or at negative 
$m$, thus not invariant under $Z_2$. On the contrary, the eigenvalues for $\epsilon\geq \epsilon_*$ 
are not degenerate and are $Z_2$ symmetric. More specifically, any eigenfunction 
$\Phi_\epsilon(m)$ has evanescent tails that vanish exponentially with $N$ in the regions where 
$f_-(m) > \epsilon$ and $f_+(m) < \epsilon$. In Fig. \ref{evan} we show in the case 
$\epsilon<\epsilon_*$ the regions of evanescent waves. In this case, one can construct two 
eigenfunctions, each localized in a well, whose mutual overlap 
vanishes exponentially for $N\to\infty$. 
This result also implies that the ground state, which lies in the $k=0$ subspace, spontaneously breaks 
$Z_2$ when $h<h(0)$, hence $h(0)=2=h_c$ is the critical transverse field at which the quantum phase transition takes place. Such a degenerate ground state is actually a wave packet centered either at 
$m=+\sqrt{1-h^2/4}$ or at $-\sqrt{1-h^2/4}$, see Eq.~\eqn{m-eq}.

More generally, it follows that for any given $k$ and $h<h(k)=2(1-2k)$ there is indeed a "broken-symmetry edge" 
\be
\epsilon_*(k) = -h\,(1-2k),\label{mobility-edge}
\ee
that separates symmetry breaking eigenstates at $\epsilon<\epsilon_*$ from symmetric eigenstates at 
higher energies. In particular, in the lowest energy subspace with $k=0$, the edge is 
$\epsilon_*(0)=-h$. Therefore, although in the simple mean-field like model Eq.~\eqn{Ham}, one can 
indeed prove the existence of energy edges that separate symmetry invariant from symmetry breaking eigenstates. We also note that subspaces corresponding to different $k$ have different $\epsilon_*(k)$, as we anticipated in the Introduction.

\subsection{The role of the broken-symmetry edge in the quench dynamics}

\begin{figure}[ht]
\centerline{\includegraphics[width=8cm]{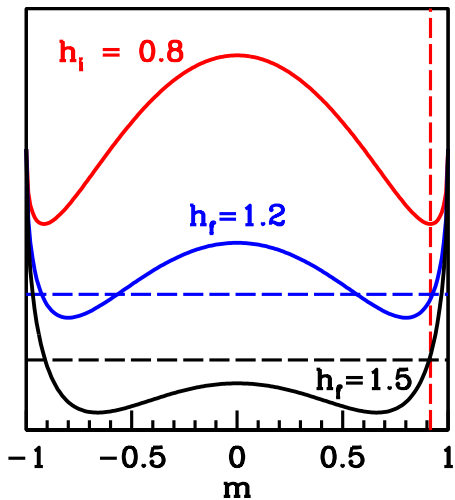}}
\vspace{-0.6cm}
\caption{\label{quench} (Color online) Pictorial view of the quench dynamics. The upper curve (red) corresponds to the Hamiltonian for $t<0$ characterized by $h_i=0.8$. We assume that for negative time the wave function is the ground state for $m>0$, which corresponds to the minimum of the curve at positive $m_i$, dashed vertical line. The intermediate curve (blue) corresponds to $f_-(m)$ for $h_f=1.2$, while the lowest curve (black) to $h_f=1.5$. 
The intercepts $\epsilon=f_-(m_i)$ define the lower bound of allowed energies.}
\end{figure}

The quench dynamics we examine corresponds to propagating the ground state at $h=h_i$ with a different transverse field 
$h=h_f>h_i$. In the specific case of a fully-connected model, this problem has been addressed by Refs. \onlinecite{Sengupta} and \onlinecite{SciollaBiroli_long}. In particular, it has been found\cite{SciollaBiroli_long} 
that for $h_i<h_c$, i.e. starting from the broken-symmetry phase, a dynamical transition occurs at 
$h_f=h_* = (h_c+h_i)/2$. For $h_f\geq h_*$, the symmetry is dynamically restored, while, below, it remains broken as in the initial state. 
If, instead of a sudden increase from $h_i<h_c$ to $h_f$, one considers a linear ramp 
\[
h(t) = \begin{cases} 
h_i + (h_f-h_i)\,\fract{t}{\tau} & \text{for } t\in[0,\tau],\\
h_f & \text{for } t>\tau,
\end{cases}
\]
then the critical $h_*$ increases and tends asymptotically to the equilibrium critical value $h_c(0)$ 
for $\tau\to\infty$,\cite{Sandri} as expected for an adiabatically slow switching rate. This result demonstrates that the dynamical transition 
is very much the same as the equilibrium one, and it occurs for 
lower fields $h$ only because of anti-adiabatic effects, which is physically plausible. 

The dynamical transition can be easily discovered in the semiclassical limit. If we set 
\ba
\langle S^z\rangle &=& \frac{N}{2}\,\cos\theta,\\
\langle S^x\rangle &=& \frac{N}{2}\,\sin\theta\,\cos\phi,\\
\langle S^y\rangle &=& \frac{N}{2}\,\sin\theta\,\sin\phi,
\ea
for large $N$ it is safe to assume $\langle S^z S^z\rangle \sim \langle S^z\rangle\langle S^z\rangle = 
N^2\cos^2\theta/4$, so that the total energy per spin, see Eq. \eqn{Ham}, reads 
\be
e = \fract{\langle \mathcal{H}\rangle}{N} = 
-\cos^2\theta - h\,\sin\theta\,\cos\phi,\label{e-semi}
\ee
and is conserved in the unitary evolution. Through the Heisenberg equation $i\dot{S}^z = [S^z,\mathcal{H}]$ and upon expressing $\cos\phi$ as function of $e$ and $\theta$ by means of Eq. \eqn{e-semi}, one finds the equation of motion of the order parameter $m=\cos\theta$
\be
\fract{\partial \cos\theta}{\partial t} = 
\mp 2\sqrt{\left(h\sin\theta-e-\cos^2\theta\right)\left(h\sin\theta+e+\cos^2\theta\right)}.
\ee
It follows that, until $e<-h$, the order parameter oscillates around a finite value -- the symmetry remains broken -- while for $e>-h$ it oscillates around zero -- the symmetry is restored. The dynamical transition thus occurs when $e_*=-h$, which we recognize to be the edge between symmetry-breaking and symmetry-invariant eigenstates at $k=0$ previously defined.    

This correspondence can be shown to hold also in the exact solution of the previous section. To this end, imagine to start from the ground state at an initial $h_i<h_c$, which occurs in the subspace of $k=0$, i.e. maximum total spin $S=N/2$, and let it evolve with the Hamiltonian at a different $h_f>h_i$. We note that, since $S$  is conserved, the time-evolved wave function will stay in the subspace $k=0$.
The ground state at $h_i$ is degenerate, and we choose the state with 
a positive average of $m$, which is a wave-packet narrowly centered around  $m_i = \sqrt{1-(h_i/2)^2}$. 
For very large $N$, when the contributions from the evanescent waves in the regions where $\Re \text{e} \phi_\epsilon(m) >0$ can be safely neglected, the initial wave function decomposes in $k=0$ eigenstates of the final Hamiltonian with eigenvalues $\epsilon$ such that $f_-(m_i)\leq \epsilon \leq f_+(m_i)$. In Fig. \ref{quench} we show graphically the condition $\epsilon\geq f_-(m_i)$ for $h_i=0.8$ and two values of $h_f=1.2, 1.5$. We note that the minimum value of the allowed energy for $h_f=1.2$ belongs to the subspace of symmetry broken states, while for $h_f=1.5$ it belongs to the subspace of symmetric states. It follows that, while for $h_f=1.2$ the long time average of $m$ will stay finite, for $h_f=1.5$ it will vanish instead. The critical $h_f=h_*$ is such that $f_-(m_i) = f_-(0) = -h_*$, namely $h_*=1+h_i/2 = (h_c+h_i)/2$, 
which is indeed the result of Ref. \onlinecite{SciollaBiroli_long}. Therefore, the dynamical restoration of the symmetry is intimately connected to the existence of an energy threshold. When the initial wave function decomposes into eigenstates of the final Hamiltonian that all have energies higher than that threshold, 
then the long time average of the order parameter vanishes although being initially finite. 

We note that the dynamical transition in this particular example is related to the equilibrium quantum phase transition, 
but it is actually unrelated to the transition at finite temperature.\cite{Semerjian}
In fact, at a given value of the transverse field $h<h_c$, all eigenstates within the subspaces with $k\geq h_c -h$ are symmetric, while those with smaller $k$ have still low-energy symmetry-breaking eigenstates. Since the degeneracy $g(S)$, see Eq. \eqn{g(S)}, increases exponentially in $N$ upon lowering $S$, hence raising $k$, the entropic contribution of the symmetric subspaces at large $k$ will dominate the free energy and eventually drive the finite temperature phase transition.  On the contrary, the quench-dynamics is constrained within the subspace at $k=0$, hence it remains unaware that in other subspaces the eigenstates at the same energy 
are symmetric. 

This observation is important and makes one wonders how the above result can survive beyond the fully-connected limit. Indeed, as soon as the Fourier transform of the exchange $J_\bq$, see Eq. \eqn{Ham-0}, acquires finite components at $\bq\not=\bnot$, states with different total spin, hence different $k$, start to be coupled one to each other -- the total spin ceases to be a good quantum number, the only remaining one being the total momentum. Therefore,  symmetry-breaking eigenstates at low $k$ get coupled to symmetric eigenstates at large $k$. In this more general situation, there are to our knowledge no rigorous results apart from the pathological case of one-dimension, where the energy above which symmetry is restored actually coincides with the ground state energy, or, more rigorously, where excited states that breaks the symmetry 
do not have extensive entropy. However, we mention that a recent attempt to include 
small $\bq\not=\bnot$  fluctuations on top of the results above, i.e. treating 
\be
-\frac{1}{N}\sum_{\bq\not=\bnot}\, J_\bq\,\sigma^z_\bq \sigma^z_{-\bq} 
\label{perturb}
\ee
as a small perturbation of the {\sl bare} Hamiltonian
\[
\mathcal{H}_0 = -\frac{J_\bnot}{N}\,\sigma^z_\bnot \sigma^z_{\bnot} 
- h\,\sigma^x_{\bnot},
\]
suggests that the dynamical transition in the quench does survive,\cite{Sandri} which we take as an indirect evidence that the energy edge does, too. We suspect the reason being that, once symmetry-breaking and 
symmetry-invariant eigenstates of same energy get coupled by \eqn{perturb}, the new eigenstates, being linear combinations of the former ones, will all be symmetry-breaking. As a result, the 
broken-symmetry edge will end to coincide approximately with 
$\epsilon_*(k=0)$, i.e. with the maximum value among all the formerly independent subspaces.     

\section{A different example: the Hubbard model} \label{Hubbard}
In the Introduction we inferred the existence of a threshold energy by the existence of a finite temperature phase transition that ends at $T=0$ into a quantum critical point. However, we found in the previous section an energy edge above which symmetry is restored that is actually unrelated to the finite temperature phase transition, while it is only  linked to the quantum phase transition. This suggests that such an edge 
could exists in a broader class of situations. In some simple cases, one can actually prove its existence without much effort. Let us consider for instance the fermionic Hubbard model in three dimensions, with 
Hamiltonian 
\be
\mathcal{H} = -\sum_{i,j,\sigma}\, t_{ij}\,\Big(c^\dagger_{i\sigma}c^\dagga_{j\sigma} + H.c.\Big)
+ U\sum_i\, n_{i\up}n_{i\down},\label{Hubb}
\ee
where $U>0$ is the on-site repulsion. In the parameter space where magnetism can be discarded, the low energy part of the spectrum is that of a normal metal, hence all eigenstates are expected to be invariant under the symmetries of the Hamiltonian $\mathcal{H}$, namely translations, spin-rotations and gauge transformations. We observe that the high-energy sector of the many-body spectrum obviously corresponds to the low energy spectrum of the Hamiltonian $-\mathcal{H}$. The latter is characterized by an opposite band dispersion but, more importantly, by an attractive rather than repulsive interaction. As a result of the Cooper instability, the low-energy spectrum of $-\mathcal{H}$ must comprise eigenstates $\mid \Psi\rangle$ with superconducting off-diagonal long range order, 
\be
\lim_{|i-j|\to\infty} \langle \Psi\mid d^\dagger_{i\up}d^\dagger_{i\down}\,
d^\dagga_{j\down}d^\dagga_{j\up}\mid\Psi\rangle = \mid\Delta\mid^2>0,
\label{ODLRO}
\ee
which are not invariant under gauge transformations. Since these are identically the high-energy eigenstates of the original repulsive Hamiltonian Eq. \eqn{Hubb}, it follows that the upper part of its many-body spectrum 
contains eigenstates with off-diagonal long range order. There must thus exist a special energy which separates low-energy gauge-symmetric eigenstates from high-energy superconducting ones. 

Suppose to prepare a wave function that initially has $\mid\Delta(t=0)\mid^2>0$, see Eq. \eqn{ODLRO},  
and let it evolve with the Hamiltonian $\mathcal{H}$, Eq. \eqn{Hubb}.  If its energy is high enough,  so 
that its overlap with the upper part of the spectrum is finite, then $\mid\Delta(t\to\infty)\mid^2>0$, otherwise 
$\mid\Delta(t\to\infty)\mid^2\to 0$, signaling once again a dynamical transition that, unlike the previous 
example of section \ref{Ising}, is now accompanied by the emergence rather than disappearance of long-range order. 
This surprising result was already conjectured by Rosch {\sl et al.} in Ref. \onlinecite{Rosch} as a possible metastable state attained by initially preparing a high energy wave function with all sites either doubly occupied or empty at very large $U$. What we have shown here is that such a superfluid behavior is robust and it is merely a consequence of the high energy spectrum of $\mathcal{H}$, which contains genuinely superconducting eigenstates. We finally observe that these states have negative temperature, hence are invisible in thermodynamics unless one could effectively invert the thermal population.\cite{Werner-U}

\begin{figure}[hbt]
\vspace{0.6cm}
\centerline{\includegraphics[width=7.cm]{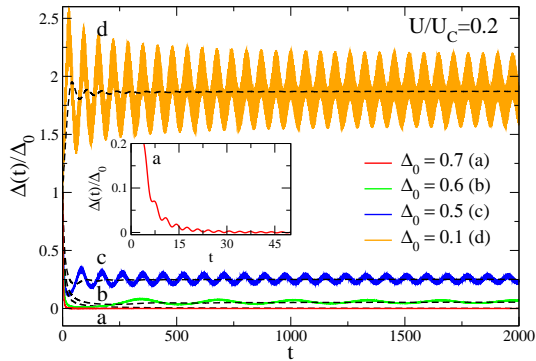}}
\caption{Time evolution of the superconducting order parameter $\Delta(t)$ for a repulsive $U=0.2 U_c$, where $U_c$ is the critical repulsion at the Mott transition, and four 
different initial values $\Delta_0$, the curves a, b, c and d. We also show in the inset the early time relaxation of $\Delta(t)$ in case a. 
Time is is units of the inverse of half the bandwidth. We observe that the curves b, c and d maintain a finite order parameter, unlike the curve a, although it corresponds to the largest initial $\Delta_0=0.7$. We also note that in the case d with the lowest $\Delta_0=0.1$, the order  parameter actually grows in time.  } 
\label{Delta}
\end{figure}

While the existence of a high-energy subspace of superfluid eigenstates of the Hamiltonian \eqn{Hubb} is evident by the above discussion, it is worth showing explicitly its consequences in the 
out-of-equilibrium dynamics. To this end, we prepare an initial wave function $\mid \Psi(t=0)\rangle =\mid \Psi_\lambda\rangle$, ground state of the BCS Hamiltonian 
\be
\mathcal{H}_\text{BCS} = +\sum_{i,j,\sigma}\, t_{ij}\,\Big(c^\dagger_{i\sigma}c^\dagga_{j\sigma} + H.c.\Big)
+ \sum_i\, \Big(\lambda\,c^\dagger_{i\up}c^\dagger_{i\down} + H.c.\Big),\label{BCS}
\ee
where the sign of the hopping is changed with respect to \eqn{Hubb}, and $\lambda$ is a control parameter 
that allows to span the alleged high-energy superconducting subspace of the repulsive Hubbard model 
by tuning the value of the initial order parameter  
\be
\Delta_0 \equiv
\Delta(t=0) = 
\langle 
\Psi_{\lambda}
\mid 
c^\dagger_{i\up}c^\dagger_{i\down} +
c_{i\down} c_{i\up}  
\mid \Psi_{\lambda}\rangle.
\label{order-parameter}
\ee

Since this model proved insoluble 
so far, we resort to an approximate method for simulating the unitary time-evolution: the 
time-dependent Gutzwiller approximation.\cite{SchiroFabrizio_short,SchiroFabrizio_long}  In brief, the time-evolved wave function is approximated by the form 
\be
\mid\Psi(t)\rangle \simeq \prod_i\,\mathcal{P}_i(t)\,\mid\Psi_{\lambda(t)}\rangle,
\ee
where $\mathcal{P}_i(t)$ is a non-hermitian time-dependent variational operator 
that acts on the Hilbert space 
at site $i$, and $\mid\Psi_{\lambda(t)}\rangle$ the solution of the Schr{\oe}dinger equation with a BCS Hamiltonian like in Eq. \eqn{BCS} but with time-dependent coupling constants $t_{ij}$ and $\lambda$. 
\begin{figure}[hbt]
\vspace{-0.4cm}
\centerline{\includegraphics[width=7cm,angle=-90]{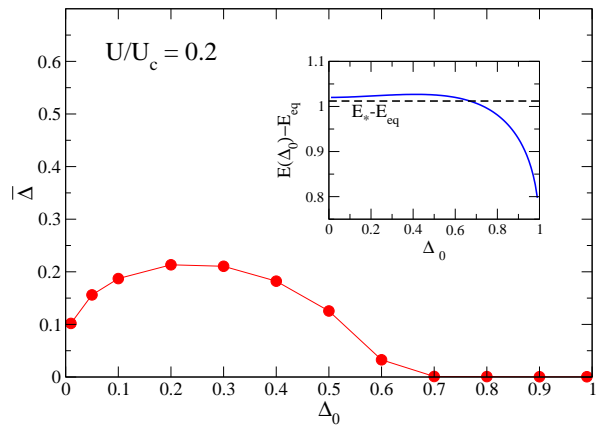}}
\vspace{-.5cm}
\caption{Long-time average of the order parameter, $\bar{\Delta}$, as function of its initial value, $\Delta_0$. In the inset we show the energy, in units of half-bandwidth, with respect to the ground state one obtained within the Gutzwiller approximation,  as function of $\Delta_0$, as well as the broken-symmetry edge, the dashed line, 
corresponding to $U=0.2 U_c$. We note that the energy, i.e. the average value of the Hamiltonian 
\eqn{Hubb} on the ground state of \eqn{BCS}, is lower the greater $\Delta_0$ because of the sign change 
of the hopping in the two Hamiltonians.} 
\label{Delta-av}
\end{figure}
All variational parameters are determined through the saddle point of the action 
\be
\mathcal{S} = \int dt\, \langle\Psi(t)\mid i\frac{\partial}{\partial t} - \mathcal{H}\mid\Psi(t)\rangle,
\label{action}
\ee
which is computed within the Gutzwiller approximation. The method has been described in detail elsewhere, see for instance Ref. \onlinecite{Hvar},  
and we just present the results here. For convenience, all calculations are done at half-filling, where they are much simpler, focusing on the paramagnetic sector, i.e. discarding magnetism, and assuming a flat density of states
with half-bandwidth $D$ that we take as the energy unit. 
However, in order to avoid spurious interference effects from the Mott localization that occurs above a critical $U_c$, we concentrate on values of $U$ safely below $U_c$, 
in fact below the dynamical analogue of the Mott transition that is found when $U\gtrsim U_c/2$.
\cite{WernerPRL,SchiroFabrizio_short}

In fig. \ref{Delta} we plot the time evolution of the order parameter 
for $U=0.2 U_c$ and different initial values of $\Delta_0$, 
see Eq. \eqn{order-parameter}, as well as different energies, see inset of Fig. \ref{Delta-av}, where we also show the time-averaged values of $\Delta(t)$ with respect to their initial values. 
We observe that at the lower energies, which actually correspond to the larger $\Delta_0$, the order parameter, initially finite, relaxes rapidly to zero. On the contrary, above a threshold energy, $\Delta(t)$ stays finite and even grows with respect to its initial value, see Fig. \ref{Delta-av}. If we identify that threshold energy as our broken-symmetry edge, although rigorously it is only a lower bound, its dependence upon $U$ is shown in Fig. \ref{mobility-edge}.
\begin{figure}[hbt]
\vspace{0.6cm}
\centerline{\includegraphics[width=8cm]{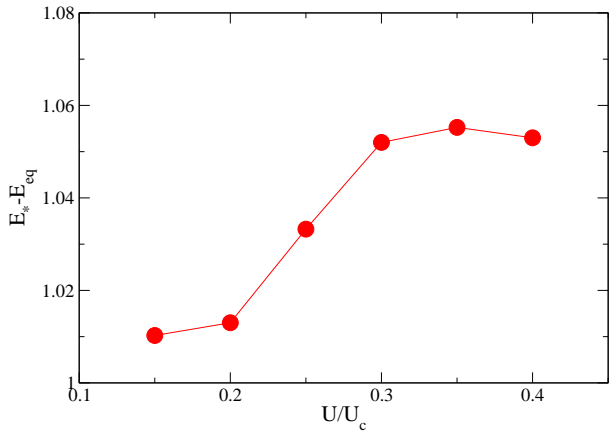}}
\caption{Broken-symmetry edge measured with respect to the ground state energy for different values of $U$} 
\label{mobility-edge}
\end{figure}

The fermionic Hubbard model that we have discussed so far it is only a very simple example where one can infer the existence of a threshold energy in the many-body spectrum that is not accompanied by any anomaly in the thermodynamics. In fact, the idea of comparing the low-energy spectrum of a Hamiltonian $\mathcal{H}$ with that of $-\mathcal{H}$, which is actually the high-energy spectrum of $\mathcal{H}$, is very simple yet very effective. However, we must remark that model Hamiltonians $\mathcal{H}$, like the Hubbard model above, are meant to describe low energy properties of complex physical systems. Therefore, it is not unlikely that the value of the threshold energies extracted by comparing the low energy spectra of $\mathcal{H}$ and $-\mathcal{H}$ could be above the limit of applicability of the model itself, in which case they would be devoid of physical relevance.

\section{Conclusions}

The original purpose of this work was to understand how an isolated macroscopic quantum system brought away from equilibrium by a sudden injection of energy could cross a order-disorder phase transition as if its temperature were raised.  We argued that this is possible because the eigenvalue spectrum of the system is characterized by a well defined energy, which we named {\sl broken-symmetry edge}, that separates low energy symmetry-breaking eigenstates from high-energy symmetry-invariant ones. Once the initial energy exceeds such a threshold, a dynamical restoration of symmetry occurs that mimics the equilibrium phase transition as obtained by increasing temperature or adiabatically changing coupling constants. We explicitly demonstrated this 
mechanism in the fully-connected Ising model in a transverse field, a prototypical mean-field example of a quantum phase transition. 

After that, we realized that such edges might arise also when the equilibrium phase diagram does not 
display any order-disorder phase transition, and bring about unexpected dynamical features. 
For instance, the repulsive Hubbard model has a high energy sector of genuinely superfluid eigenstates. By simulating in 
the Hubbard model the unitary evolution of a BCS wave function we surprisingly found that, if the energy exceeds a threshold, the initial superfluid order parameter not only survived in the long time limit, but may actually grew, despite the fact that the interaction is repulsive.

In summary we believe that the concept of broken-symmetry edges that separate well distinct sectors of the eigenvalue spectrum 
is quite general and may be a useful framework of 
reference in the out-of-equilibrium quantum dynamics of 
isolated systems.

\section*{Acknowledgments}
We thank Erio Tosatti, Claudio Castellani and Marco Schir\`o for discussions and comments. 
This work has been supported by 
the European Union, Seventh Framework Programme, under the project
GO FAST, grant agreement no. 280555.


\end{document}